\documentclass[prb,showpacs,twocolumn,superscriptaddress, a4paper]{revtex4}
\usepackage{graphicx,epsfig}
\usepackage{amsmath}
\usepackage{bm}
\usepackage{color}
\usepackage{hyperref}
\newcommand{\mri}{\mathrm{i}}

\begin{document}
\title{Attractive Hubbard Model on a Honeycomb Lattice}

\author{K.L.~Lee}
\affiliation{Centre for Quantum Technologies, National University of Singapore; 2 Science Drive 3 Singapore 117542}
\affiliation{NUS Graduate School for Integrative Sciences and Engineering, National University of Singapore, Singapore}
\affiliation{Laboratoire Kastler Brossel, UPMC-Paris 6, ENS, CNRS; 4 Place Jussieu,F-75005 Paris, France}
\author{K.~Bouadim}
\affiliation{Department of Physics, Ohio State University; 191 West Woodruff Ave Columbus OH 43210-1117, USA}
\author{G.G.~\surname{Batrouni}}
\affiliation{INLN, Universit\'e de Nice-Sophia Antipolis, CNRS; 1361 route des Lucioles, 06560 Valbonne, France}
\affiliation{Centre for Quantum Technologies,
National University of Singapore; 2 Science Drive 3 Singapore 117542}
\author{F.~H\'ebert}
\affiliation{INLN, Universit\'e de Nice-Sophia Antipolis, CNRS; 1361 route des Lucioles, 06560 Valbonne, France}
\author{R.T. Scalettar}
\affiliation{Physics Department, University of California, Davis, California 95616}
\author{C.~Miniatura}
\affiliation{INLN, Universit\'e de Nice-Sophia Antipolis, CNRS; 1361 route des Lucioles, 06560 Valbonne, France}
\affiliation{Centre for Quantum Technologies,
National University of Singapore; 2 Science Drive 3 Singapore 117542}
\affiliation{Department of Physics, National University of Singapore, %
2 Science Drive 3, Singapore 117542, Singapore}
\author{B.~Gr\'emaud}
\affiliation{Laboratoire Kastler Brossel, UPMC-Paris 6, ENS, CNRS; 4 Place Jussieu,F-75005 Paris, France}
\affiliation{Centre for Quantum Technologies, National University of Singapore; 2 Science Drive 3 Singapore 117542}
\affiliation{Department of Physics, National University of Singapore, %
2 Science Drive 3, Singapore 117542, Singapore}

\date{\today}

\begin{abstract}
We study the attractive fermionic Hubbard model on a honeycomb lattice using
determinantal quantum Monte Carlo simulations. By increasing the interaction strength
$U$ (relative to the hopping parameter $t$) at half-filling and zero temperature, the
system undergoes a quantum phase transition at $5.0<U_c/t<5.1$ from
a semi-metal to a phase displaying simultaneously superfluid behavior and
density order. Doping away from half-filling, and increasing the interaction strength at finite but low
temperature $T$, the system always appears to be a superfluid exhibiting a
crossover between a BCS and a molecular regime. These different regimes are analyzed by studying the
spectral function. The formation of pairs and the emergence of phase
coherence throughout the sample are studied as $U$ is increased and $T$ is lowered.
\end{abstract}

\pacs{03.75.Ss,	05.30.Fk, 71.10.Fd, 71.30.+h, 71.10.Pm}

\maketitle

The recent discovery of graphene layers, {\it i.e.} single-atom thick
layers of carbon atoms arranged in a planar honeycomb
structure,\cite{novoselov2004} has attracted considerable attention due
to its interest in fundamental physics as well as for potential
applications. The energy band spectrum shows ``conical points" where the valence and conduction bands are connected, and
the Fermi energy at half-filling is located precisely at these points as
only half of the available states are filled.  Around these points, the
energy varies proportionally to the modulus of the wave-vector and the excitations
(holes or particles) of the system are equivalent to ultra-relativistic (massless) Dirac
fermions since their dispersion relation is linear.\cite{semenoff1984} Graphene sheets then allows for table-top
experiments on two-dimensional field theories with quantum anomalies,
allowing us to explore the Klein paradox,\cite{katnelson2006} the  anomalous quantum Hall
effect induced by Berry phases\cite{novoselov2005,zhang2005} and its
corresponding modified Landau levels.\cite{li2007}

When the fermions are interacting, the peculiar nature of the Fermi surface ({\it i.e.} reduced to a finite
number of Dirac points) leads to special physics at and
around half-filling.  In a square
lattice, the nesting of the Fermi surface generally leads to ordered
phases even for arbitrarily small interaction strengths.  On the
contrary, in the honeycomb lattice and with repulsive interactions,
Paiva {\it et al.} have found\cite{paiva2005} a quantum
phase transition (QPT) at half-filling between a metallic and an ordered phase when the interaction strength is increased.  However, since graphene is a
weakly-interacting system, this QPT is not accessible experimentally.

In a recent work, some of us have analyzed the possibility of reproducing graphene physics and of extending it to the interacting regime by creating a two-dimensional honeycomb optical lattice and loading ultracold spin-1/2 fermionic atoms, such as $^6$Li, into it.\cite{lee2009} The key advantage is that the relevant experimental parameters ({\it e.g.} configuration and strength of the optical potential, inter-atomic interaction strength tuned via Feshbach resonance) can be accurately controlled while getting rid of the inherent complexity of a solid. Following this idea, we use exact Quantum Monte Carlo (QMC) simulations to study interacting ultracold fermions loaded into a honeycomb optical lattice in the absence of any external confinement. We will focus on the case of attractive interactions as it is accessible with these numerical techniques and free from the sign problem at and away from half-filling.

In the continuum at zero temperature, as the interacting
fermionic gas is driven from the weak to the strong attractive coupling limit,
there is a crossover from a BCS regime of weakly-bound delocalized pairs
to a Bose-Einstein condensate (BEC) of tightly-bound pairs (later called
molecules for simplicity).\cite{leggett1980,nozieres1985,randeria1994} At finite
but sufficiently low temperature, a similar BCS-molecule crossover is observed except that, the system being two-dimensional, there is only quasi-long-range order and, consequently, no true condensate but only a
superfluid. In this paper, we will study interacting particles on a
lattice, represented by a simple fermionic Hubbard model.\cite{hubbard1967} Nonetheless, some aspects of the continuum limit, such as the BCS-BEC crossover,
are expected to be reproduced in the discrete model. Zhao and
Paramekanti have explored the attractive fermionic Hubbard model on a
honeycomb lattice using mean field theory \cite{zhao2006} and they found a
QPT between a semi-metal and a superfluid at
half-filling. Away from half-filling, they recovered the crossover already
observed in the continuum limit.  Recently, Su {\it et al.} used QMC methods to study the BCS-BEC crossover on the honeycomb lattice away from half-filling and concluded that it was similar to the one obtained for the square lattice.\cite{su2009} In the present work, we use QMC simulations and large system sizes to study the pair formation at half-filling and accurately determine the critical value of the coupling strength at which pairs form. We then study pairing away from half-filling by analyzing several quantities, including spectral functions.

The paper is organized as follows. In section I, we introduce the model, notations and the quantities we use to characterize the different phases. In
section II, we show that our system at half-filling can be related to
the repulsive Hubbard model\cite{paiva2005} and then present
complementary results for this case, including the QPT point  the system crosses to go from a semi-metallic disordered phase to an
ordered one displaying both superfluid behavior and density wave order. The location of this QPT point has been accurately determined
compared to previous works, and the nature of the weakly-interacting phase
before the transition is addressed by analyzing the behavior of the spectral function as the interaction strength is varied. Finally,
in section III we study the system doped away from half-filling. The
system is clearly shown to exhibit superfluid behavior while the density wave order present at half-filling has been destroyed. We conclude our study by analyzing the formation of pairs
and the emergence of global phase coherence as a function of temperature and
interaction strength.

\section{The fermionic Hubbard model}
The physics of a system of $N_{\rm f}$ spin-1/2 fermions, with attractive two-body interactions and equal spin populations, filling up a lattice made of $N$ sites is encapsulated in a simple tight-binding model, namely the fermionic attractive Hubbard model (FAHM), whose grand-canonical Hamiltonian operator reads:\cite{paiva2004}
\begin{eqnarray} \label{model}
H &=& -t \sum_{\langle i,j\rangle, \sigma} 
\left( f^\dagger_{i\sigma} f_{j\sigma} + f^\dagger_{j\sigma} f_{i\sigma}  \right)\\
&\,& -U \sum_i \left(n_{i\uparrow} -1/2\right) \left(n_{i\downarrow} -1/2\right)
- \mu \sum_{i,\sigma} n_{i\sigma}. \nonumber
\end{eqnarray}
Here $\langle i,j\rangle$ denotes pairs of
nearest-neighbors sites on the lattice, $\sigma = \uparrow,
\downarrow$ are the two possible spin states of the fermions, $f^\dagger_{i\sigma}$ and $f_{i\sigma}$ are the creation and annihilation operators of a fermion with spin state $\sigma$ at site $i$, $n_{i\sigma}=f^\dagger_{i\sigma} f_{i\sigma}$ is the corresponding number operator, $t$ is the hopping amplitude between nearest-neighbors sites, $U\geq 0$ is the strength of the attractive interaction between fermions with opposite spin states and $\mu$ is the chemical potential whose value fixes the average total fermionic density $\rho$. With the present form of the interaction term, the system is half-filled, {\it i.e.} there is on average one fermion per site ($\rho = N_{\rm f} / N =1$), when $\mu =0$. In the non-interacting limit $U=0$, this system is known to behave like a semi-metal with vanishing density of
states at the Fermi level and its elementary excitations are massless Dirac
fermions that obey the 2D Weyl-Dirac equation.\cite{neto2009}

The FAHM (\ref{model}) on a bipartite lattice is particle-hole
symmetric\cite{lieb1989} and thus adopts the same phases for densities $\rho$
and $2-\rho$. It is then sufficient to study the system for densities
$\rho \ge 1$.  This model can also be mapped onto the fermionic repulsive Hubbard
model (FRHM)\cite{hubbard1967,paiva2005} by performing a particle-hole
transformation on only one of the species. Consequently, the physics of the FAHM at densities $(\rho_\uparrow, \rho_\downarrow)$ is equivalent to that of the FRHM at densities $(1-\rho_\uparrow, \rho_\downarrow)$ or
$(\rho_\uparrow, 1-\rho_\downarrow)$, but with a non-zero Zeeman-like term, $-\mu\sum_{i}\left(n_{i\uparrow}-n_{i\downarrow}\right)$. Therefore, the two models are identical at half-filling ($\mu=0$). We will use this equivalence
in section \ref{sec:halffilling} where we concentrate on the half-filled
case.

\begin{figure}[!ht]
\includegraphics[width=0.47\textwidth]{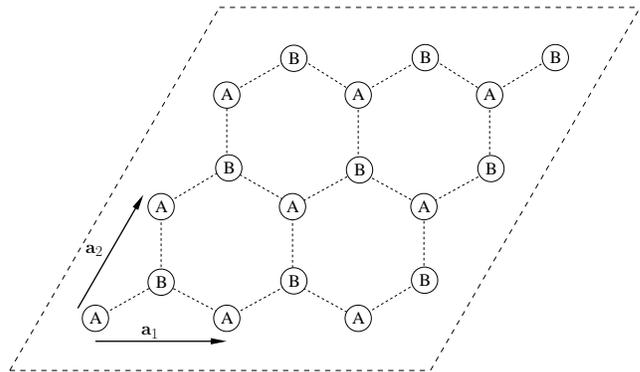}
\caption{\label{fig:lattice}Finite honeycomb lattice of linear
  dimension $L= 3$. The total number of sites is $N=2L^2=18$.}
\end{figure}

To calculate the equilibrium properties of this model at finite but low
temperatures $T$, we used the
standard determinant quantum Monte Carlo algorithm (DQMC).
\cite{hirsch1983, white1989, scalettar1989, loh1990, santos2003} The
cases under our consideration (namely attractive interactions and equal
densities of spin-up and spin-down fermions) are free of the sign
problem\cite{loh1990} that used to plague numerical simulations of
fermionic systems. This will allow us to reach the low temperatures needed to study pairing and superfluidity. In the following, the reciprocal of the thermal energy (also called the inverse temperature) is denoted as usual by $\beta=1/k_BT$, where $k_B$ is the Boltzmann constant.

\begin{figure}[!ht]
\includegraphics[height=0.47\textwidth,angle=-90,bb=53 290 300 610]{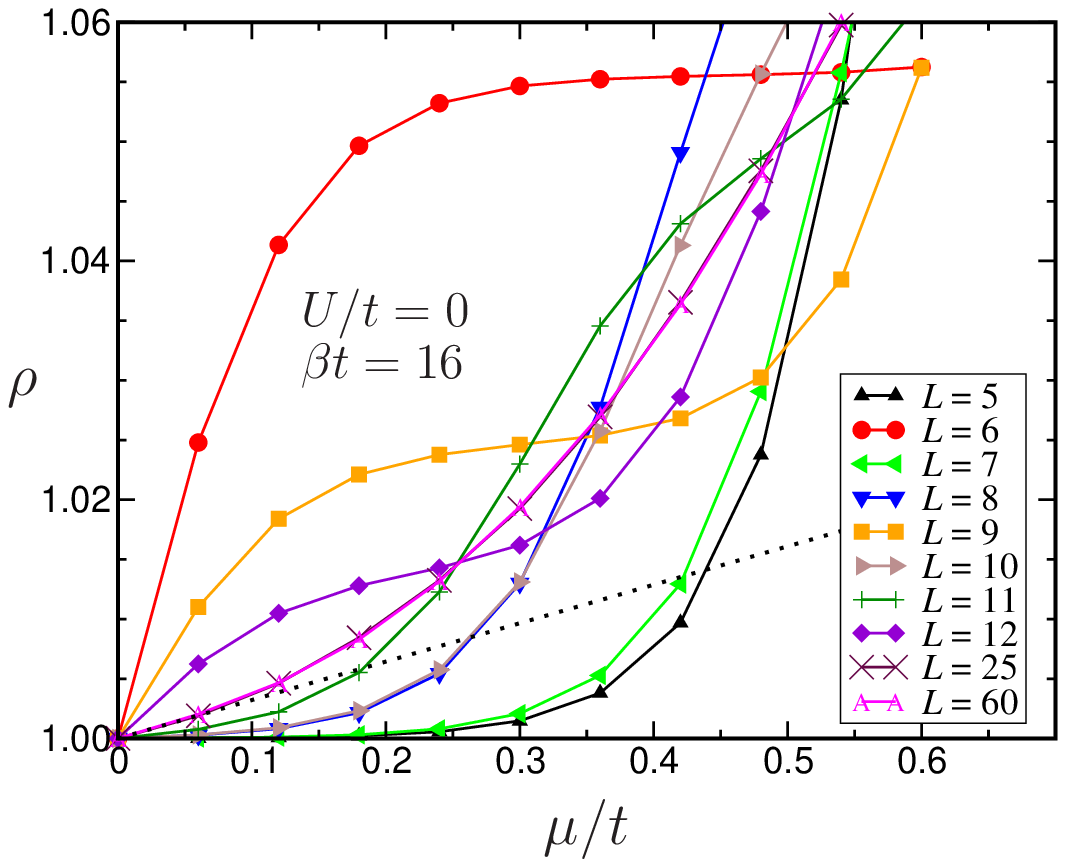}
\includegraphics[height=0.47\textwidth,angle=-90,bb=53 290 300 610]{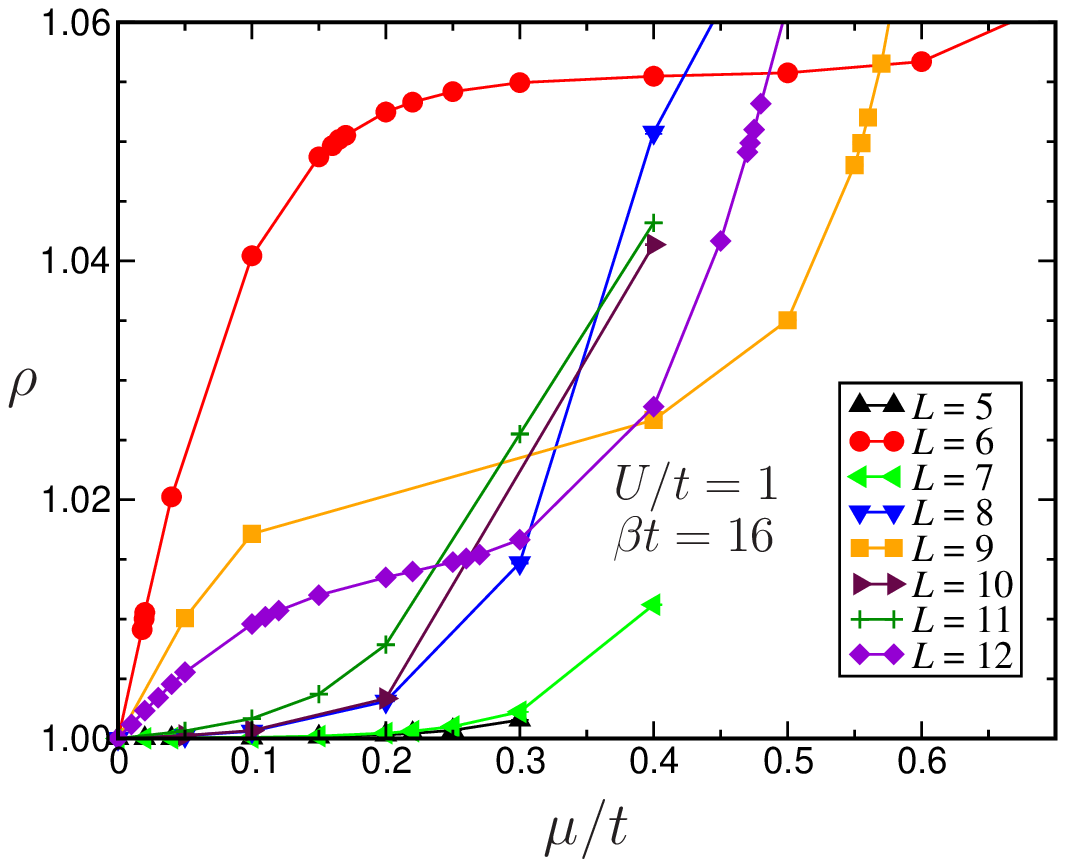}
\caption{\label{fig:densityvsmu}(Color online) Total average density $\rho$ {\it vs}
chemical potential $\mu$ for $U/t = 0$ (top) and $U/t = 1$ (bottom) at
$\beta t=16$ and different lattice sizes $L$. The top figure is obtained by
analytical calculation at $U=0$. The bottom figure
is obtained from numerical data generated by DQMC. For sizes that are
not multiples of three, there is no state at half-filling and a small gap
appears for small system sizes. There is no such gap when $L$ is a
multiple of three. For sizes that are multiples of three, plateaus appear away from half-filling. These plateaus are also finite-size effects and they disappear when $L\to\infty$. The dotted line in the top figure is obtained by an exact
evaluation of the derivative $\partial \rho/\partial \mu|_{\mu=0}$ in the
non-interacting limit when $L\to\infty$. The two figures show
that the "magic number 3" effect is present even when the interaction strength
$U$ is comparable to the hopping parameter $t$.}
\end{figure}

In the DQMC simulations, we have used the honeycomb lattice depicted in Fig.
\ref{fig:lattice} with periodic boundary conditions. The primitive vectors ${\bf a}_1$
and ${\bf a}_2$ delineate a diamond-shaped primitive cell of the Bravais lattice which
contains two nonequivalent sites (\textsc{a} and \textsc{b}) separated by 
$\overrightarrow{{\rm \textsc{ab}}} =({\bf a}_1 + {\bf a}_2) /3$ and each producing upon tiling a hexagonal sublattice.
A finite honeycomb lattice of side $L$ then contains $N= 2L^2$
sites.  In the non-interacting case, the energy levels are given by
\cite{semenoff1984,lee2009} $$ \epsilon_\pm (k_1, k_2) = \pm t \left| 1
+ e^{i 2\pi k_1 / L} + e^{i 2\pi k_2 / L}\right|,$$ where $k_1, k_2 \in
\{0, 1, \, \cdots \,, L-1\}$. When $L$ is a multiple of three, there always exist pairs 
$(k_1,k_2)$ such that $\epsilon_\pm (k_1,k_2) = 0$, {\it i.e.} there are
four states (two per spin state) located exactly at the Fermi level and only
two of these states will be occupied if $\rho=1$. This does not happen
when $L$ is not a multiple of three. As a consequence, on small finite-size
systems, a small gap of order $1/L$ appears around half-filling when $L$
is not a multiple of three (see Fig. \ref{fig:densityvsmu}).  To avoid
confusion between this gap, which is a finite-size effect, and Mott gaps
generated by interactions that are expected to appear in
ordered phases, we used (especially at half-filling) sizes $L$ that are
multiples of three. This limits strongly the sizes that can be studied.  In
the most favorable cases, we went up to $L=15$, that is $N=450$ sites.

In the strong coupling regime ($U\gg t$), we expect the system to form
pairs (hereafter called molecules) of fermions with opposite spins on the same site. These
pairs can show two different ordering phenomena: establishment of a
phase coherence order or of a solid (crystal-type) order. A solid of pairs would exhibit 
a density wave typical of a crystal and would reveal itself through spatial oscillations in
the density-density correlation function,
\begin{equation}
D_{ij} = \langle n_i n_j\rangle,
\end{equation}
where $n_i= \sum_{\sigma} n_{i\sigma}$ is the total number of fermions on site $i$ and where $\langle \cdot\rangle$ denotes the quantum statistical average at temperature $T$. At
half-filling and zero temperature, we expect to observe a phase where alternate sites are
empty and where only the \textsc{a} or the \textsc{b} sub-lattice is
occupied. Such a density wave is signaled by a structure factor $S_{\rm
dw}$ diverging linearly with the total number of sites $N$ of the
system, where
\begin{equation}
S_{\rm dw} = \frac{1}{N} \sum_{i,j} (-1)^{i+j} \, D_{ij} 
\end{equation}
with the site index $i$ being even on \textsc{a} sites and odd on
\textsc{b} sites.

In a Bose condensed phase, the phase coherence
between pairs is signaled by long-range order (or quasi-long-range order
for a superfluid at finite temperature) in the pair Green's function,
\begin{equation}
G^{\rm p}_{ij} = \frac{1}{2} \langle \Delta^\dagger_i \Delta_j + \Delta_i \Delta^\dagger_j \rangle,\label{eqn:pairgreen}
\end{equation}
where $\Delta^\dagger_i = f^\dagger_{i\uparrow}f^\dagger_{i\downarrow}$
creates a pair on site $i$. In a way similar to the density
correlations, we define a pair structure factor $P_{\rm s}$,\footnote{The s index
indicates the symmetry of the wave function, by analogy with the
notation of the hydrogen orbitals. Here, the on-site pair is invariant
by rotation.}
\begin{equation}
P_{\rm s} = \frac{1}{N} \sum_{i,j} G^{\rm p}_{ij}.
\end{equation}
This pair structure factor diverges linearly with N when long-range order is achieved. Finally, in the absence of any order, the system is expected to be a semi-metal
at half-filling due to the peculiar nature of the Fermi surface (no gap
but a vanishing density of states at the Fermi level). To distinguish between metallic, 
semi-metallic or gapped (solid or superfluid) states, we calculate the
spectral function $A(\omega)$ which essentially reflects the one-particle density of
states. To obtain this quantity, we first calculate the
(imaginary) time-displaced on-site Green's function $ G(\tau) =\sum_i
\langle f_i(\tau) f_i^\dagger(0) \rangle /N $ and then extract $A(\omega)$ by
inverting the following Laplace transform 
$$
G(\tau) = \int d\omega
\frac{e^{-\tau \omega}}{e^{-\beta \omega} + 1} A(\omega)
$$ 
using an analytic continuation method.\cite{sandvik1998}

\section{Honeycomb lattice at half-filling\label{sec:halffilling}}
At half-filling, the system can be mapped onto the
FRHM.\cite{robaszkiewicz1981a,robaszkiewicz1981b,robaszkiewicz1981c,moreo1991}
Defining a hole creation operator $h^\dagger_{i\downarrow}$ for the down
spin through,
\begin{equation}
(-1)^i h^\dagger_{i\downarrow} = f_{i\downarrow},
\end{equation}
the kinetic term is left unchanged in the spin-down holes
representation. The number operator $n_{i\downarrow}$ is accordingly transformed
into $1 - n^h_{i\downarrow}$, where $n^h_{i\downarrow} =
h^\dagger_{i\downarrow} h_{i\downarrow}$ is the number operator for holes, and, up to a redefinition of the chemical potential $\mu$, the sign of the
interaction term is reversed. The FRHM has SU(2) spin-rotation symmetry
at half-filling, which translates into the SU(2) pseudo-spin symmetry of
FAHM.\cite{yang1991} Hence the spin-spin correlations are the same along
the three coordinate axes,
\begin{equation}
\langle \sigma_i^x \sigma_j^x\rangle = \langle \sigma_i^y\sigma_j^y\rangle = \langle \sigma_i^z\sigma_j^z\rangle, \label{eq:spincorrel}
\end{equation}
where $x$ and $y$ are the in-plane axes and $z$ the axis orthogonal to the lattice plane. More specifically:
\begin{eqnarray}\label{pauli}
	\sigma_i^x &=& f^\dagger_{i\uparrow}h_{i\downarrow}+h^\dagger_{i\downarrow}f_{i\uparrow},\\
	\sigma_i^y &=& \mri(h^\dagger_{i\downarrow}f_{i\uparrow}-f^\dagger_{i\uparrow}h_{i\downarrow})\nonumber \\
	\sigma_i^z &=& n_{i\uparrow}-n_{i\downarrow}^h. \nonumber
\end{eqnarray}
At large interaction, the FRHM is known to be equivalent to a Heisenberg
model and it develops a long-range anti-ferromagnetic order on the
honeycomb lattice at zero temperature.\cite{paiva2005} The correlation
functions (\ref{eq:spincorrel}) then show oscillations from site to
site. Translated into the attractive model language, these functions
become \cite{zhang1990,lieb1989}
\begin{eqnarray}
\label{correlPauli1} \langle \sigma_i^z \sigma_j^z\rangle &=& \langle n_i n_j - n_i - n_j - 1 \rangle,\\
\label{correlPauli2}\langle \sigma_i^x \sigma_j^x + \sigma_i^y \sigma_j^y\rangle &=&  2\,(-1)^{i+j}\,\langle \Delta^\dagger_i \Delta_j + \Delta_i \Delta^\dagger_j \rangle.
\end{eqnarray}
The spin anti-ferromagnetic correlations along the $z$-axis in
the FRHM are then reproduced in the density-density correlations $D_{ij}$
of the FAHM, which develops a density wave with alternating occupied and
empty sites. The spin correlations in the $xy$ lattice plane translate into 
long-range order for the Green's function $G^{\rm p}_{ij}$ and phase
coherence of a Bose-Einstein condensate. The anti-ferromagnetic 
phase of the FRHM is thus mapped onto a peculiar phase for the FAHM since it exhibits 
at the same time phase coherence {\em and} density wave orders. In the following we will denote this phase as the DW-SF phase. Moreover it is easy to show from equations \eqref{correlPauli1} and \eqref{correlPauli2} that $2P_{\rm s} = S_{\rm dw}$ as is numerically checked in Table \ref{tab:SandP}. As the order parameter is here of
dimension three and the lattice is of dimension two, we do not expect any transition to an ordered
phase at finite temperature.\cite{mermin1966}
\begin{table}[b!]
\renewcommand{\arraystretch}{1.3}
\begin{tabular}{|l||c|c|c||}
\cline{1-4}
\multicolumn{1}{|c||}{$\mu/t$} &\hspace{0.05cm} $\rho$ \hspace{0.05cm} & $S_{\rm dw} /2$ & $P_{\rm s}$  \\ 
\cline{1-4}
\hline
0 & 1.0 & 1.125 $\pm$  0.005 & 1.127 $\pm$ 0.001 \\
0.9202& 1.5  & 0.3356 $\pm$ 0.0004 &  10.5 $\pm$  0.1 \\
\hline
\end{tabular}
\caption{Comparison of $P_{\rm s}$ and $S_{\rm dw} / 2$ for $L=12$, $\beta t=20$,
$U/t=3$, and different values of $\mu/t$. At half-filling, those
quantities are equal within statistical error bars as a consequence of
the SU(2) pseudo-spin symmetry of the FAHM. $S_{\rm dw}$ and $P_{\rm s}$ are small because $U<U_c$ and the system is in its semi-metallic phase. This symmetry is broken when
$\mu\neq0$ and this is confirmed by the numerical data showing that the
two quantities are indeed unequal. $S_{\rm dw}$ remains small but $P_{\rm s}$ is large due to the presence of quasi-long-range order.\label{tab:SandP}}
\end{table}

\begin{figure} \includegraphics[height=0.47\textwidth,angle=-90,bb=53
245 361 640]{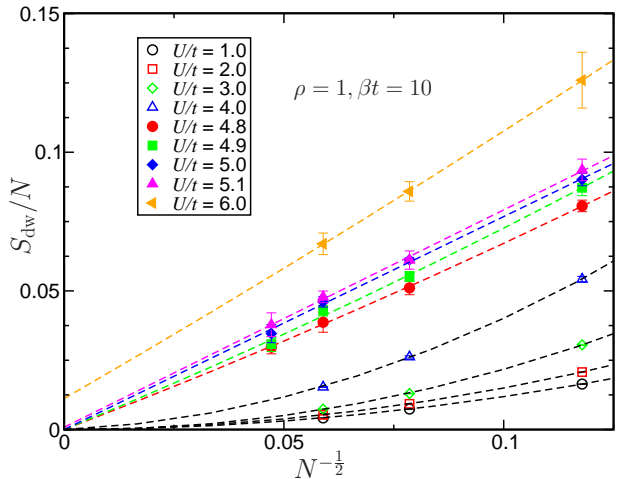}
\caption{\label{fig:scalingSHF}(color online) Scaling of the density wave structure factor $S_{\rm dw}$ with lattice size $L$ at half-filling (the total number of lattice sites is $N=2L^2$). The dashed lines are a fit of the form $S_{\rm dw}/N=a+b/\sqrt{N}+c/N$. Close to or above the transition ($U/t\gtrsim 5.0$), the coefficients $a$ and $b$ take on finite positive values implying that both density and phase coherence orders emerge in the thermodynamic limit $N\to\infty$. As it is seen, $S_{\rm dw}/N$ then essentially scales linearly with $1/\sqrt{N}$ and achieves the finite value $a$ when $N \to \infty$. Below the transition ($U/t \lesssim 5$), the coefficients $a$ and $b$ vanish, meaning that the system reaches its disordered phase in the thermodynamic limit $N\to\infty$. As it is seen, $S_{\rm dw}/N$ then essentially scales as $1/N$ and goes to zero when $N \to \infty$. The QPT point is thus signaled by the vanishing of the coefficient $a$, from which we can infer that  the critical interaction strength lie in the range $5.0<U_c/t<5.1$.}  \end{figure}

Paiva \textit{et al.}\cite{paiva2005} have studied the ground state of
FRHM on a honeycomb lattice a few years ago. They found a QPT from an anti-ferromagnetic phase at large coupling to a
metallic phase at low coupling, the critical coupling strength being bounded by
$4\leq U_c/t\leq 5$. We use finite-size scaling and larger system sizes $L$ to
improve the numerical accuracy and narrow down the region of this QPT. Spin wave theory applied to
Heisenberg models implies that the structure and pair structure factors at $T=0$ 
scale with the number of lattice sites $N=2L^2$ like \cite{auerbach1994,huse1988,
paiva2004, paiva2005} 
$$
2 P_{\rm s}(N) =  S_{\rm dw}(N)\approx a N + b\sqrt{N} +c
$$
where $a,b,c$ are $U$-dependent nonnegative constants. In the disordered phase $S_{\rm dw}(N)$ is
expected to reach a constant finite value as $N$ goes to infinity, meaning that the coefficients $a$ and $b$ should then vanish. In the ordered phase, $a$ should be strictly positive so that both $P_{\rm s}$ and $S_{\rm dw}$ diverge linearly with $N$ signaling the emergence of density and phase coherence orders. Using
system sizes as large as $L= 15$, and using the vanishing of coefficient $a$ to define the onset for the DW-SF phase, we have been able to infer the critical interaction strength $U_c$ to be
in the range $5.0 < U_c/t < 5.1$ (Fig. \ref{fig:scalingSHF}).

\begin{figure} \includegraphics[height=0.47\textwidth,angle=-90,bb=53
290 289 725]{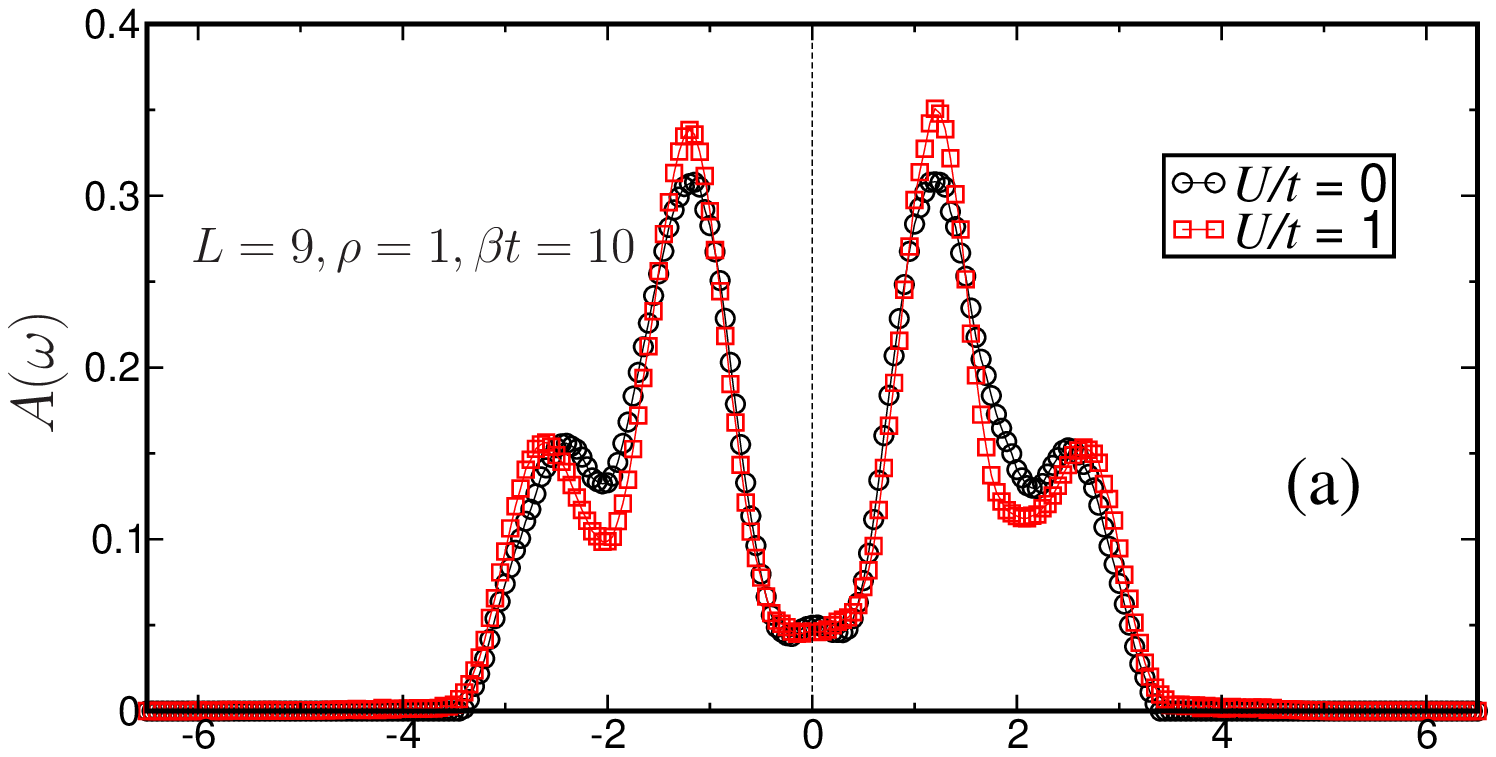}
\includegraphics[height=0.47\textwidth,angle=-90,bb=53 290 289
725]{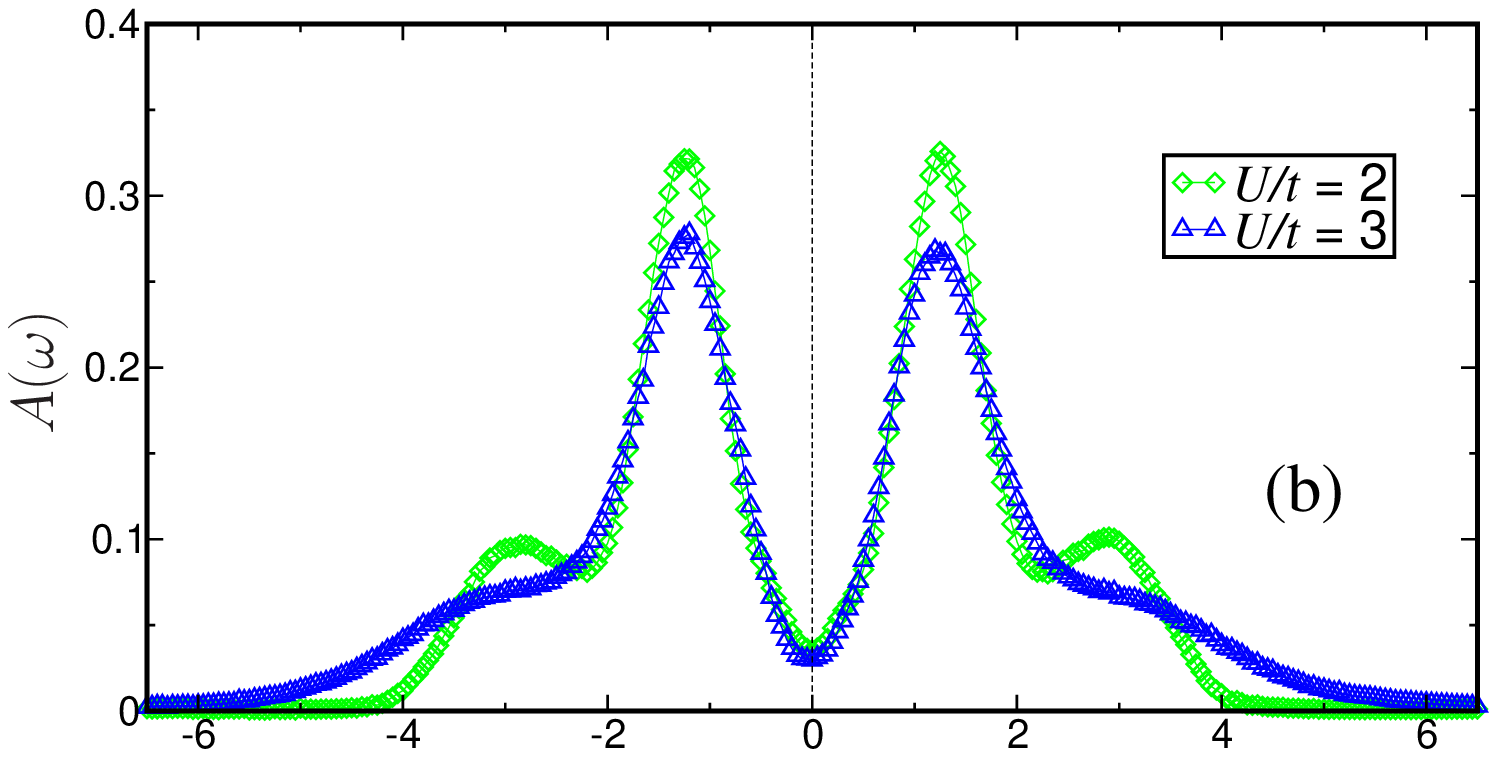}
\includegraphics[height=0.47\textwidth,angle=-90,bb=53 290 289
725]{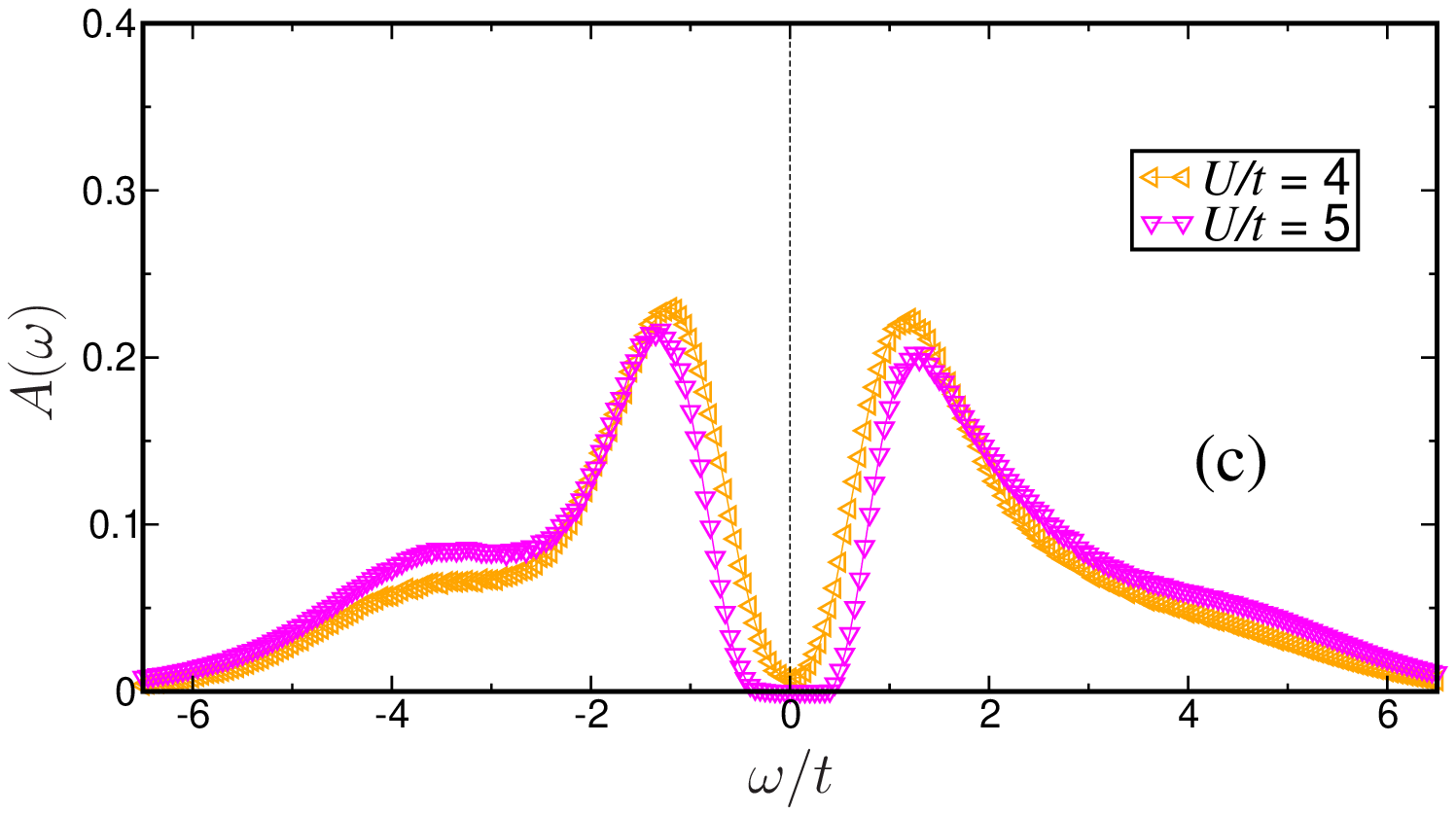}
\caption{\label{fig:spectrHF}(Color online) Spectral function
$A(\omega)$ at half-filling ($\rho=1$) for different values of the interaction strength $U$. The lattice size is $L=9$ and $\beta t =10$. The Fermi level is located at $\omega =0$. For $U/t< 5$,
the system is a semi-metal as witnessed by the dip around the Fermi level. The
non-vanishing density of states at the Fermi level is due to finite-size
effects (see Fig.~\ref{fig:densityvsmu}). For $U/t>5$, a gap opens as
the system enters the DW-SF ordered phase. The small peaks situated at
$|\omega|\approx 2.5\, t$ are also a result of finite-size effects.}
\end{figure}

In the study by Paiva \textit{et al.}, the metallic phase appearing at
low $U$ was not studied in detail. In particular the question of the
metallic or semi-metallic nature of the system was not addressed.
Calculating the spectral function $A(\omega)$ for different values of
$U$ (Fig. \ref{fig:spectrHF}), we find that the system is always a 
semi-metal when it is not in an ordered phase. The density of states drops around
the Fermi level (located at $\omega =0$) for $U/t<5$ but without forming a gap. On the
contrary, we observe a tiny metallic peak at the Fermi level. This
peak is a finite-size effect due to the four states per spin located
exactly at the Fermi level (in the non-interacting limit) when the
system size is a multiple of three. On the contrary, using sizes that are not
multiples of three, we do observe a small gap. Both this
gap and the peak are finite-size effects that are reduced when we
increase the size of the system. We then conclude that $A(\omega)$ is
zero (or very small) only at the Fermi level but without the formation
of a gap. This is the signature of a semi-metallic phase. Indeed, a metal would be
signaled by a persistent peak at the Fermi level (or at least a large
non-zero density). The transition to the DW-SF ordered phase is
signaled by the opening of the gap in $A(\omega)$ for $U/t \ge 5$,
which corresponds to the value for the transition previously obtained
by the finite-size scaling analysis of $S_{\rm dw}$.

\section{Doping away from half-filling} 

\begin{figure}[!ht]
\includegraphics[height=0.47\textwidth,angle=-90,bb=53 295 330
640]{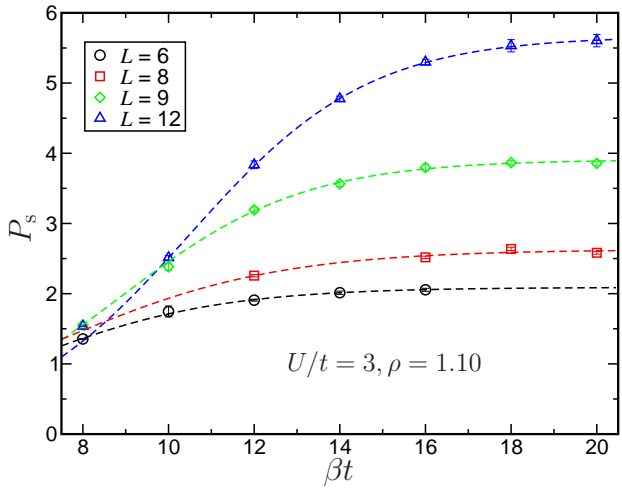} \caption{(Color online) Evolution of the pair structure
factor $P_{\rm s}$ as a function of the inverse temperature $\beta t$ for several lattice sizes $L$. The interaction strength has been fixed at $U=3t$ and the total average fermionic density at $\rho=1.1$.\label{fig:pairingvsbeta} The dashed lines are fits using the 3-parameter function $F(\beta t)$, eq.~\eqref{fit}. A plateau is reached when $\beta t$ is much greater than the energy gap induced by finite-size effects between the ground state and the first excited
state. As can be seen, the plateau is reached at larger $\beta t$ when the lattice size
increases. It is also reached at larger $\beta t$ when $\rho \to 1$ (not shown).}
\end{figure}

At zero temperature, when the FAHM is doped away from the DW-SF ordered
phase obtained at half-filling when $U>U_c$, say by increasing $\rho$ from 1, we expect the density order to disappear and the
phase coherence order to persist. However, one also expects phase coherence to establish throughout the sample when the system is doped away from the semi-metallic phase obtained at half-filling when $U < U_c$. Indeed in this case the Fermi surface is no longer
limited to isolated points and BCS pairing becomes possible. Therefore, we
expect the phase coherence order to establish at zero temperature for all
values of the interaction $U$ as soon as $\rho \not= 1$. With an order parameter of
dimension two (a phase gradient pictured as a vector lying in the
$xy$-plane), the system undergoes a Berezinskii-Kosterlitz-Thouless
(BKT)\cite{kosterlitz1973,berezinski1970,berezinski1971} transition at
some critical  temperature $T_c$, leading to a quasi-long-range phase order, {\it
i.e.}  a superfluid phase, at $T < T_c$ before the appearance
of the Bose-Einstein condensate at $T=0$.

According to mean-field theory\cite{zhao2006}, a superconductor exists
anywhere away from half-filling, albeit the superconducting gap
function or, equivalently, $\langle \Delta_i^\dagger\rangle$, decays
exponentially with respect to $1/(U\sqrt{\rho-1})$ in the BCS regime. In their previous
study\cite{su2009}, Su {\it et al.} compared DQMC results to RPA
calculations and showed that there is a so-called BCS-BEC crossover
extending from small to large values of the interaction when the system is
off half-filling. When $U$ is increased, the ground state of the system evolves continuously from a BCS state (where fermions with opposite spins form loose pairs of plane waves with opposite momenta) to a BEC of bosonic molecules (where fermions with opposite spin form tightly-bound pairs). We have extended their study to larger lattices (up to $L=15$) and lower temperatures (up to $\beta t = 20$) and we have also analyzed new observables.

We first studied the behavior of the pair and density wave structure factors, $P_{\rm s}$ and $S_{\rm dw}$, away from half-filling. To do this, we first need to obtain the low-temperature limit of
these quantities by decreasing the temperature until we observe a
plateau signaling that we have reached the $T=0$ limit (Fig.
\ref{fig:pairingvsbeta}). To extract the plateau value, we have used the 3-parameter function
\begin{equation}
\label{fit}
F(\beta t) = \frac{u}{1+v \exp(-w\beta t)}
\end{equation}
to fit our numerical data $ P_{\rm s}(\beta t)$. The plateau value $\lim_{\beta \to \infty} P_{\rm s}$ is then approximated by $u$. We have also observed in our numerical simulations that this plateau is
reached at lower and lower temperatures as we approach half-filling. This is because the BKT critical temperature $T_c$ goes to zero like $1/|\ln\delta\rho|$ as $\delta\rho =|1-\rho| \to 0$ \cite{paiva2004} and lower temperatures are required to achieve phase coherence.

\begin{figure}[!ht]
\includegraphics[height=0.47\textwidth,angle=-90,bb=53 245 350
620]{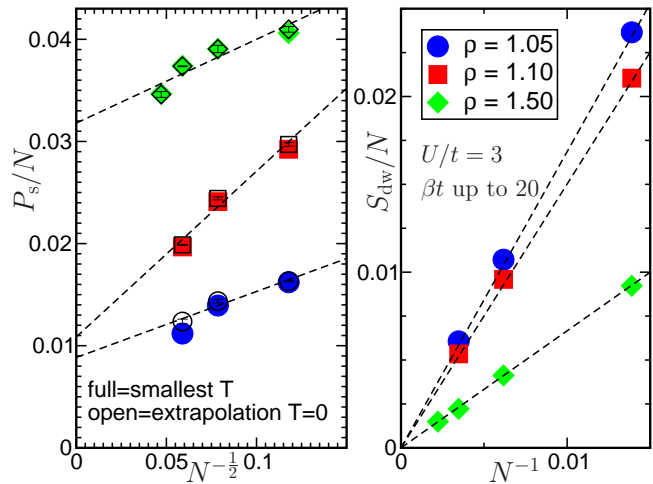} \caption{(Color online) Evolution of the pair and density wave structure
factors $P_{\rm s}$ and $S_{\rm dw}$ as a function of the number of lattice sites $N$ for different total average fermionic densities $\rho$. The interaction strength has been fixed at $U=3t$. Full symbols have been obtained for inverse temperatures up to $\beta t=20$ (see text). Open symbols for $P_{\rm s}$ are the plateau values at $T=0$ as extracted from the fits in Fig.\ref{fig:pairingvsbeta}. The density wave structure factors $S_{\rm dw}$ always go to zero as the system size $L=\sqrt{N/2}$ tends to infinity whereas the phase coherence ordering signal $P_{\rm s}$ never vanishes. The dashed lines are guides to the eyes. For the same parameters at half-filling the system would be semi-metallic and $S_{\rm dw}$ and $P_{\rm s}$ would both vanish.\label{fig:SCDWandPSvsN}} \end{figure}

Fig. \ref{fig:SCDWandPSvsN} shows how $P_{\rm
s}$ and $S_{\rm dw}$ scale with the number of lattice sites $N$. For each chosen lattice size $L$ and fermionic density $\rho$, we have run our simulations for the lowest temperature that could be numerically achieved. The temperature range that we have been able to explore was up to $\beta t =20$. As expected $S_{\rm dw}$ always goes to zero and
$P_{\rm s}$ always extrapolates to a non-zero value. We can then
conclude, from direct measurement, that the BEC at zero temperature
always appears as soon as the system is doped away from half-filling. Even
with the smallest doping that we have been studying ($\rho=1.05$, 5\%
doping), we have observed a clear persistence of the phase coherence ordering
in the large size limit.

\begin{figure}[!ht]
\includegraphics[height=0.47\textwidth,angle=-90,bb=53 255 360
630]{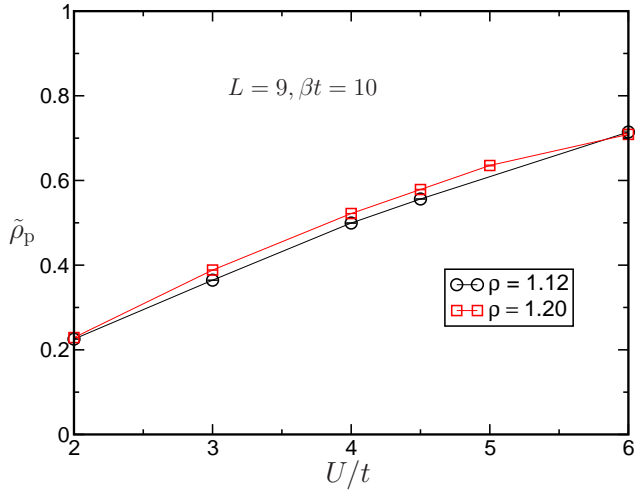} \caption{(Color online)
Evolution of the rescaled density $\tilde{\rho}_{\rm p}$ of on-site pairs, eq.~\eqref{rescaled}, as a function of the interaction strength $U/t$ for two different total average fermionic densities $\rho$. The system size has been fixed at $L=9$ and the inverse temperature is $\beta t=10$. In the non-interacting limit ($U/t\to0$), spin-up and spin-down particles are
uncorrelated, hence $\langle n_{i\uparrow} n_{i\downarrow} \rangle =
\langle n_{i\uparrow}\rangle\langle n_{i\downarrow}\rangle=\rho_\uparrow\rho_\downarrow=\rho_\uparrow^2$ for equal spin populations. In this case $\tilde{\rho}_{\rm p}=0$. In the molecular limit ($U/t\to\infty$), fermions can only exist in pair at a site, hence $\langle
n_{i\uparrow} n_{i\downarrow} \rangle=\langle n_{i\uparrow}\rangle = \rho_\uparrow$. In this case $\tilde{\rho}_{\rm p}=1$. The BCS-BEC crossover is characterized by the smooth evolution of $\tilde{\rho}_{\rm p}$ between these two limits 0 and 1 as the interaction strength is increased.\label{fig:pairdens}}
\end{figure}

To observe the molecule formation along the BCS-BEC crossover, we have studied the
density of on-site pairs
 \begin{equation} \rho_{\rm p} = \frac{1}{N}
\sum_i \langle n_{i\uparrow} n_{i\downarrow}
\rangle.\label{eqn:pairdensity} 
\end{equation} 
In the non-interacting limit ($U/t\to0$), spin-up and spin-down particles are
uncorrelated. Hence $\langle n_{i\uparrow} n_{i\downarrow} \rangle =
\langle n_{i\uparrow}\rangle\langle n_{i\downarrow}\rangle=\rho_\uparrow\rho_\downarrow$. Since we consider here equal spin populations $\rho_\uparrow=\rho_\downarrow=\rho/2$, we find $\rho_{\rm p}=\rho_\uparrow^2$. In the molecular limit ($U/t\to\infty$), fermions can only exist in pair at a site. Hence $\langle
n_{i\uparrow} n_{i\downarrow} \rangle=\langle n_{i\uparrow}\rangle = \rho_\uparrow$ and $\rho_{\rm p}=\rho_\uparrow$. In see \ref{fig:pairdens}, we have plotted the rescaled density of on-site pairs:
\begin{equation}
\label{rescaled}
\tilde{\rho}_{\rm p} = \frac{\rho_{\rm p} - \rho_\uparrow^2}{\rho_\uparrow-\rho_\uparrow^2}.
\end{equation}
as a function of $U/t$. The BCS-BEC crossover is nicely evidenced by the smooth evolution of this rescaled quantity between the two limits $\tilde{\rho}_p=0$ and $\tilde{\rho}_p=1$ as the interaction is increased.

\begin{figure}[!ht]
\includegraphics[height=0.47\textwidth,angle=-90,bb=53 290 289
725]{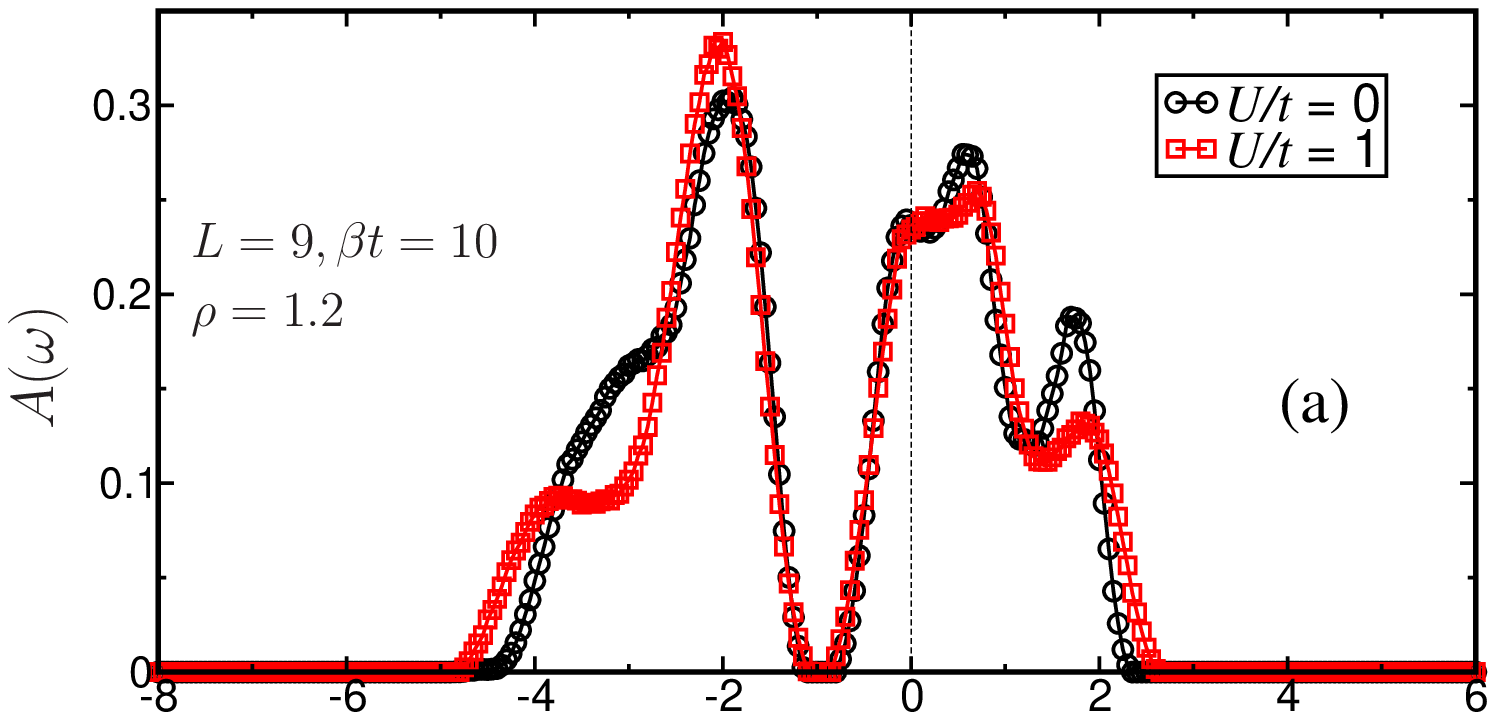}
\includegraphics[height=0.47\textwidth,angle=-90,bb=53 290 289
725]{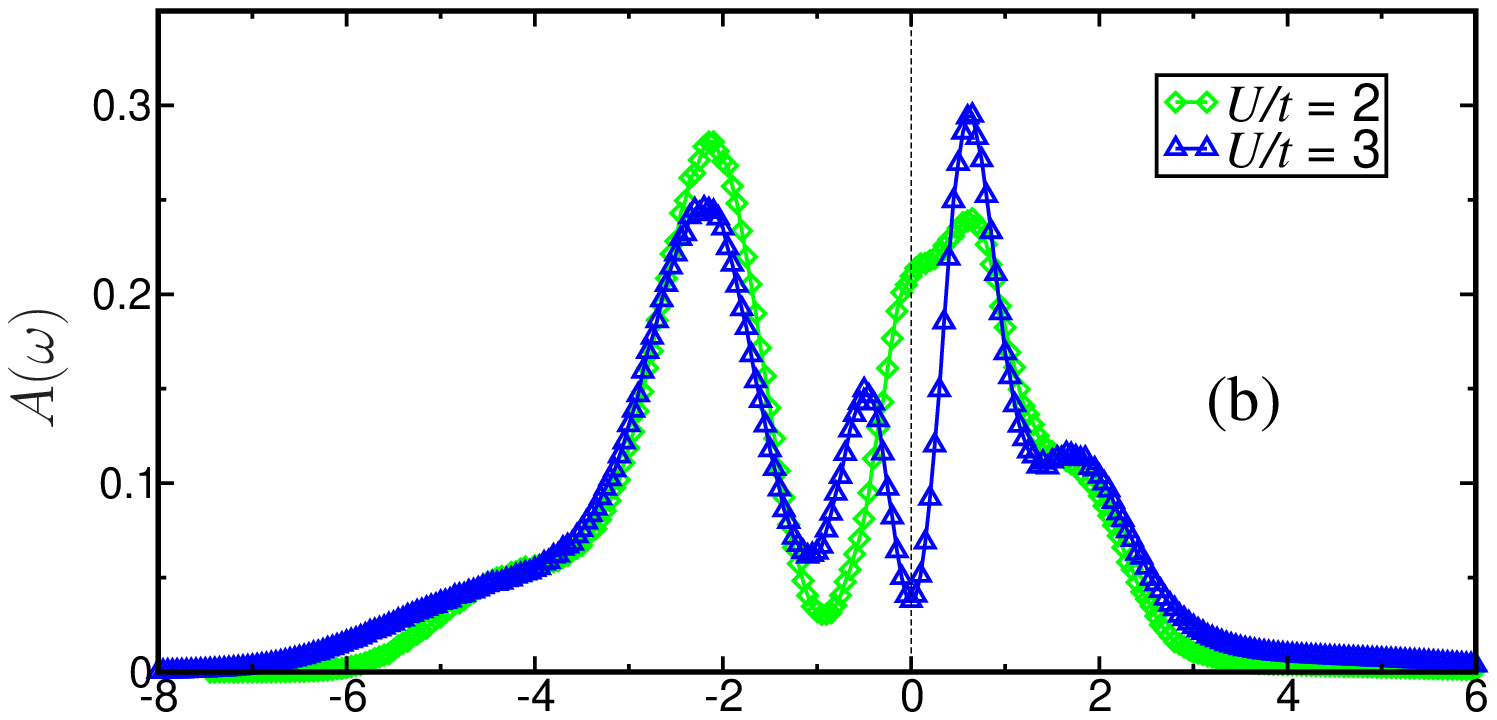}
\includegraphics[height=0.47\textwidth,angle=-90,bb=53 290 289
725]{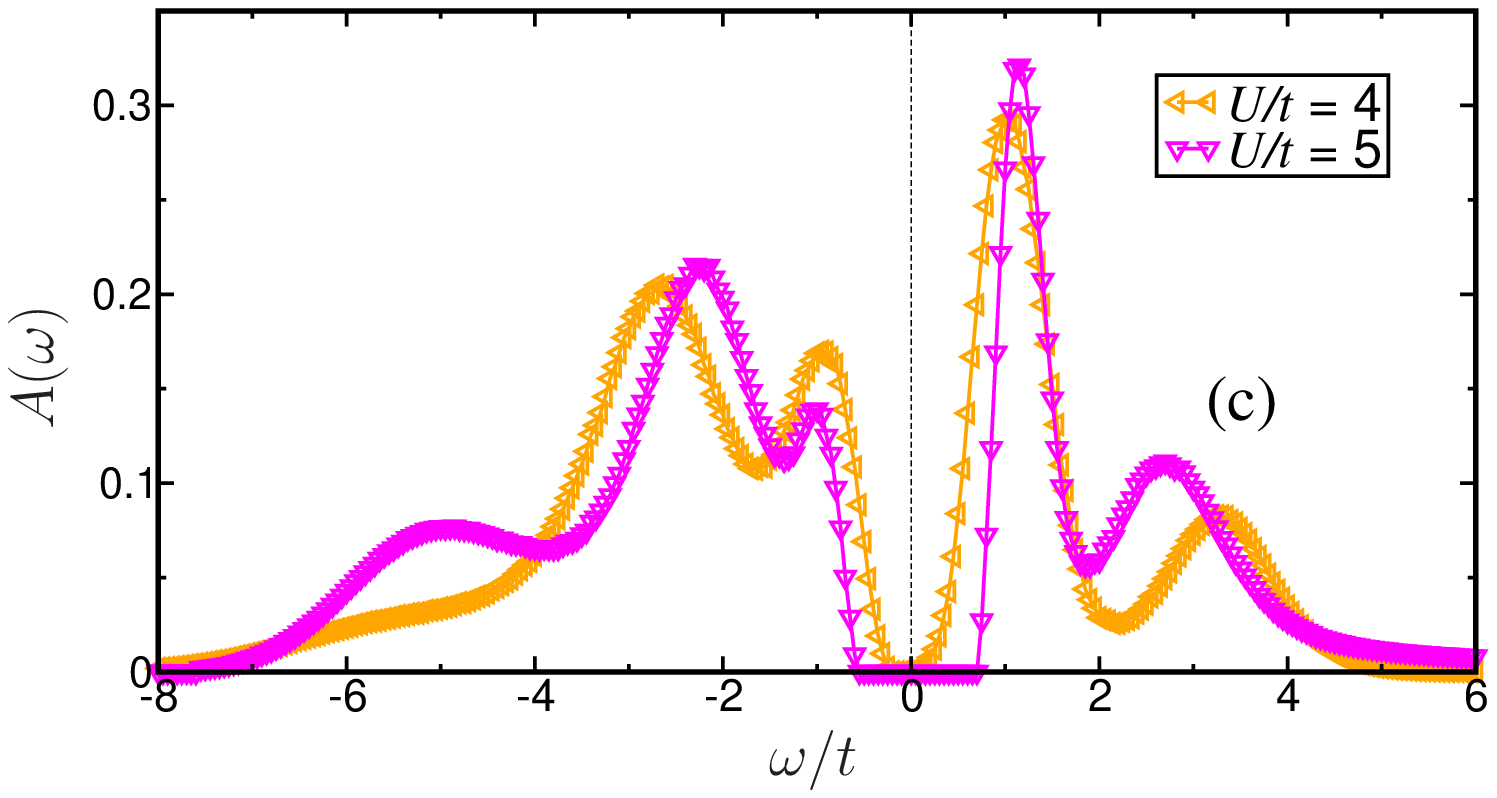} \caption{(Color online)
Evolution of the spectral function $A(\omega)$ as a function of the interaction strength $U$ at density $\rho=1.2$, inverse temperature $\beta t=12$ and lattice size $L=9$. When $U=0$, the chemical potential is numerically found to be $\mu/t=0.8768$, locating the Dirac points in the residual gap (due to finite-size effects and temperature rounding) around $\omega/t=-1$. The fact that the density of states vanishes linearly with $\omega$ around $\omega/t=-1$ also supports this identification of the location of the Dirac points. As
$U$ is increased, a dip develops in the spectral function at the Fermi level (located at $\omega = 0$) and the BCS-BEC gap eventually opens while the Dirac points are gradually destroyed. \label{fig:spectraOHF}}
\end{figure}

\begin{figure}[!ht]
\includegraphics[height=0.47\textwidth,angle=-90]{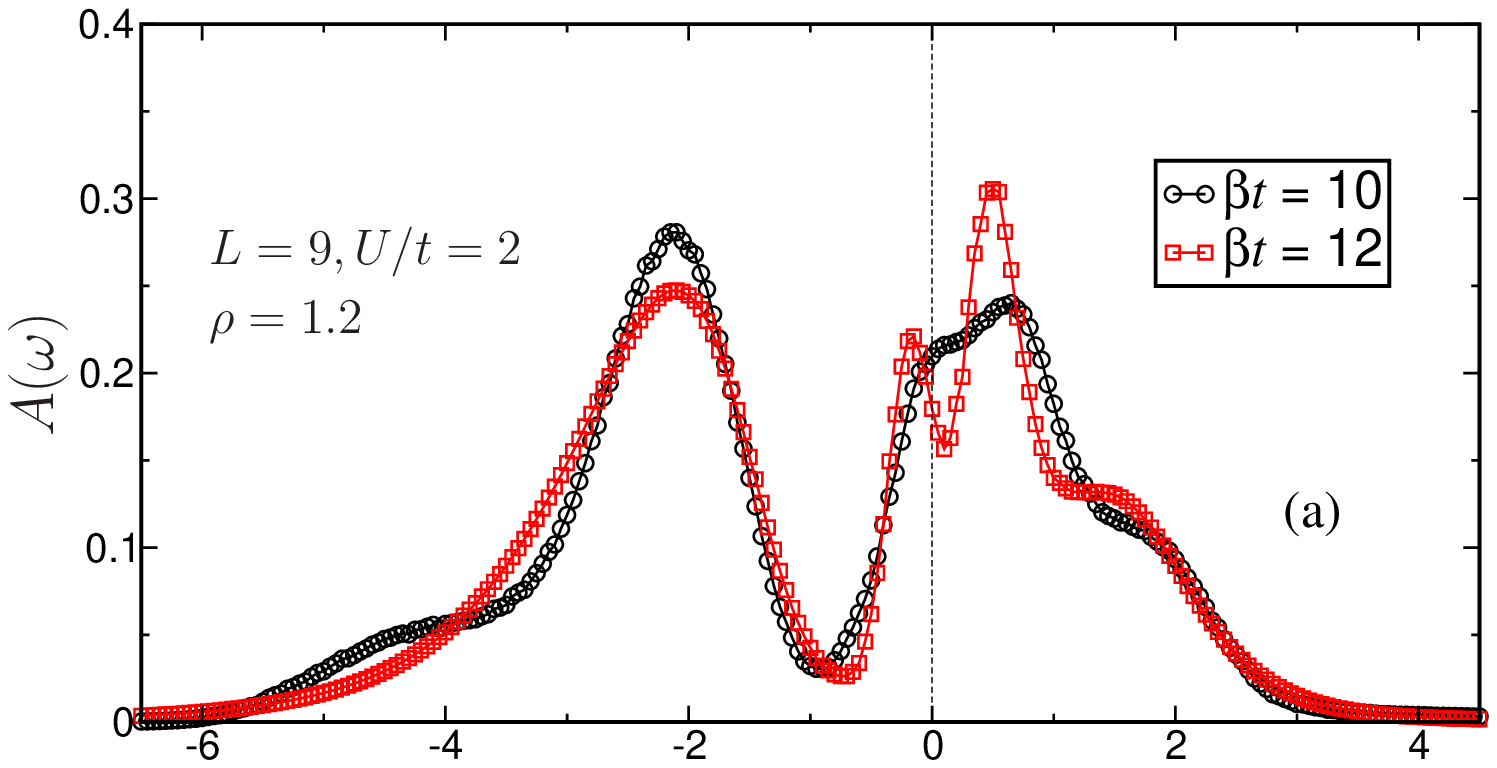}
\includegraphics[height=0.47\textwidth,angle=-90]{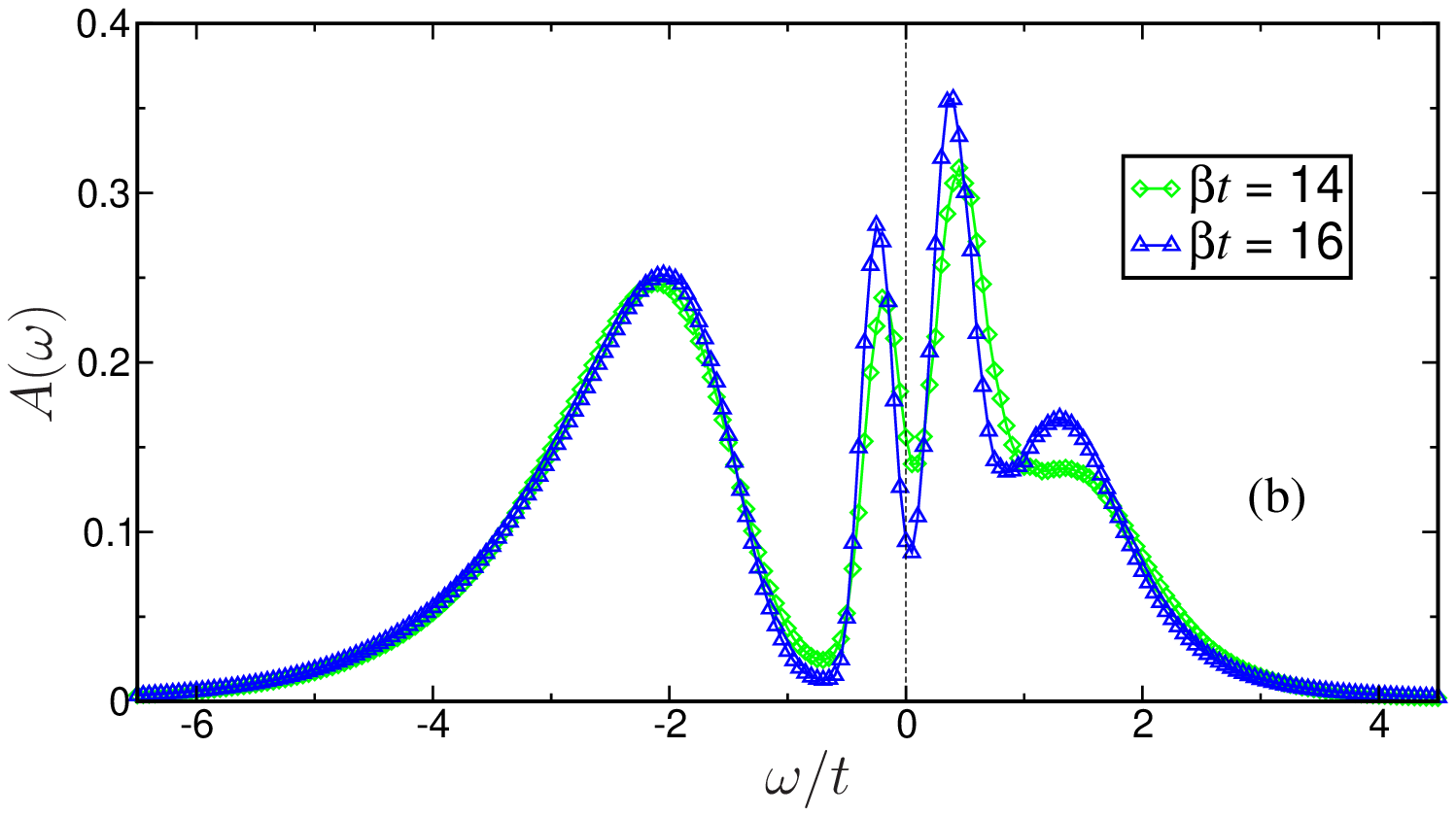}
\caption{(color online) Evolution of $A(\omega)$ as a function of inverse temperature $\beta t$ at $\rho=1.2$, interaction strength $U=2t$ and lattice size $L=9$. As the temperature is lowered, a dip develops in the spectral function at the Fermi level located at $\omega=0$. Eventually a gap opens when the temperature is low enough (not shown). The gap opening at the Fermi level is obtained even at weak interactions, a situation characteristic of the existence of a small BCS gap.  \label{fig:spectraOHFbeta}}
\end{figure}

The second evidence for molecule formation along the BEC-BCS crossover comes
from the evolution of the spectral function $A(\omega)$ when the
interaction strength $U$ (Fig.~\ref{fig:spectraOHF}) and the temperature
$T$ (Fig.~\ref{fig:spectraOHFbeta}) are varied. At large interactions ($U \ge 4$), a
clear gap is found at the Fermi level $\omega=0$ provided the temperature is low enough,
showing the formation of molecules. On the contrary, when the interaction is weaker ($U \le 3$), the gap does not open within the
same range of temperatures. However, we observe that the value of $A(\omega)$ at the
Fermi level $\omega=0$ decreases when the temperature is lowered (Fig.
\ref{fig:spectraOHFbeta}). We interpret this behavior as the precursor to
the formation of a small BCS gap at very low temperatures. This dip in
$A(\omega)$ at the Fermi level is different from the one due to the
vanishing of the non-interacting density of states at the Dirac points
that was observed at half-filling in the semi-metal case. The Dirac
dip is still present in the $U\le 3$ cases for $\omega < 0$ (Fig.
\ref{fig:spectraOHF}), showing that interaction strength is not large
enough to strongly modify the structure of the Fermi sea, except very close to
the Fermi level. This is characteristic of the BCS case. On the
other hand, the Dirac dip disappears at strong interactions
(Fig.~\ref{fig:spectraOHF}, bottom), showing now that the original Fermi sea
structure has been completely modified by interactions.

\begin{figure}[!ht]
\includegraphics[height=0.47\textwidth,angle=-90,bb=53 245 360
630]{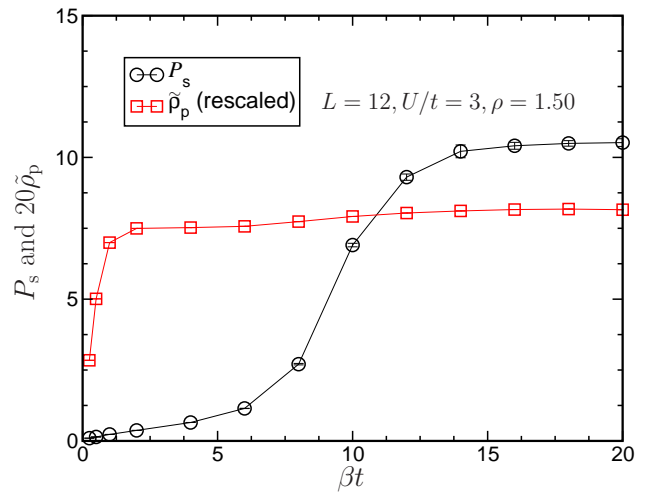} \caption{(Color online) Evolution of the pair structure factor
$P_{\rm s}$ (circles) and the rescaled density of on-site pairs $\tilde{\rho}_{\rm p}$ (squares) as a function of the inverse temperature $\beta t$ at interaction strength $U=3t$. The total average fermionic density is set at $\rho = 1.5$ and the system size is $L=12$. Two
different energy scales are clearly identified as $P_{\rm s}$, signaling the emergence of phase coherence, saturates at $\beta t\approx U/t$ whereas $\tilde{\rho}_{\rm p}$, signaling the molecule formation, saturates at $\beta t\approx t/U$. We recover here (in dimensionless units) the two energy scales $t^2/U$ and $U$, typical of the emergence of phase coherence and of the formation of tightly-bound pairs. \label{fig:Ps_and_rhop_vs_beta}}
\end{figure}

\begin{figure}[!ht]
\includegraphics[height=0.47\textwidth,angle=-90,bb=53 290 360
680]{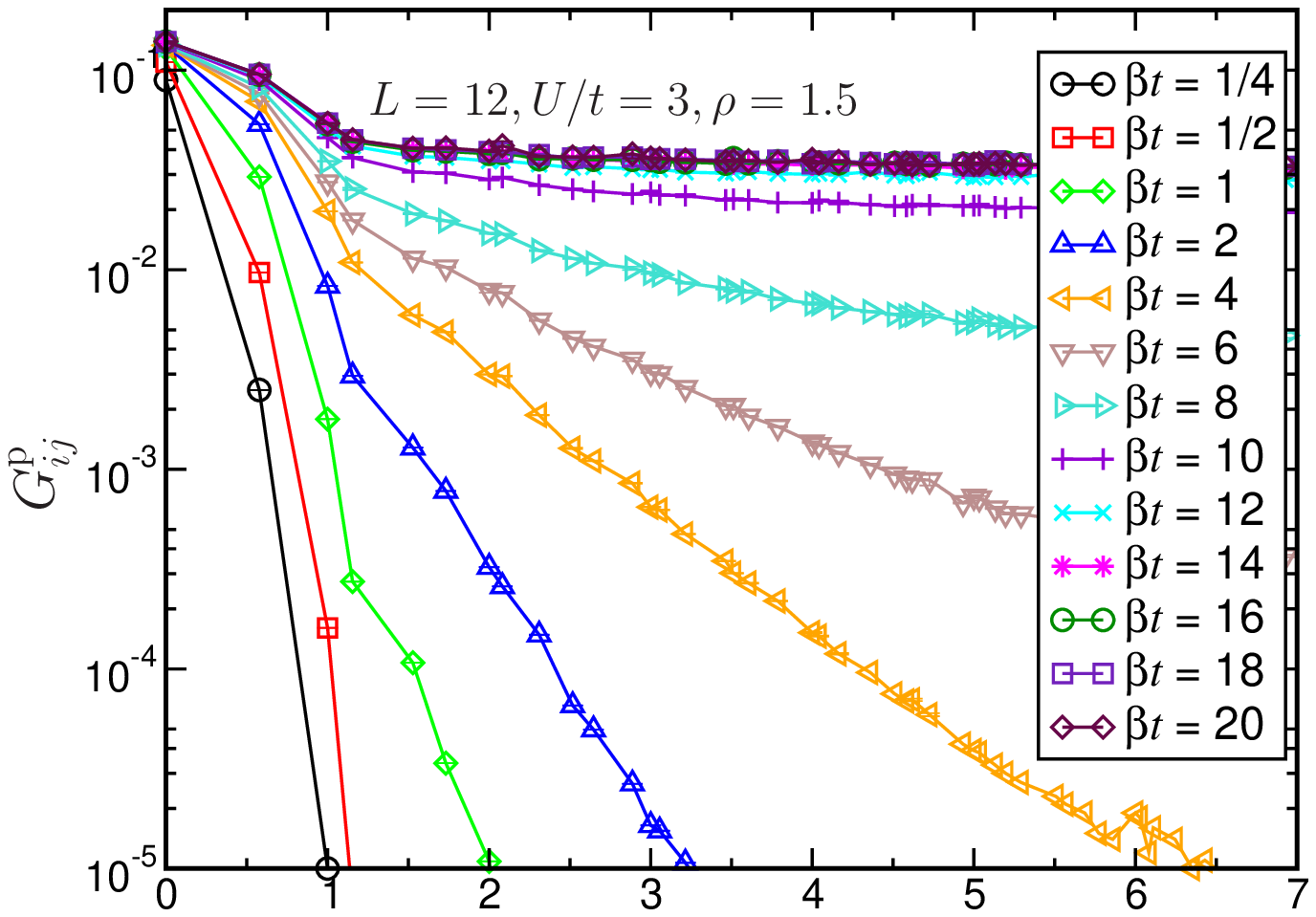}
\includegraphics[height=0.47\textwidth,angle=-90,bb=53 240 360
630]{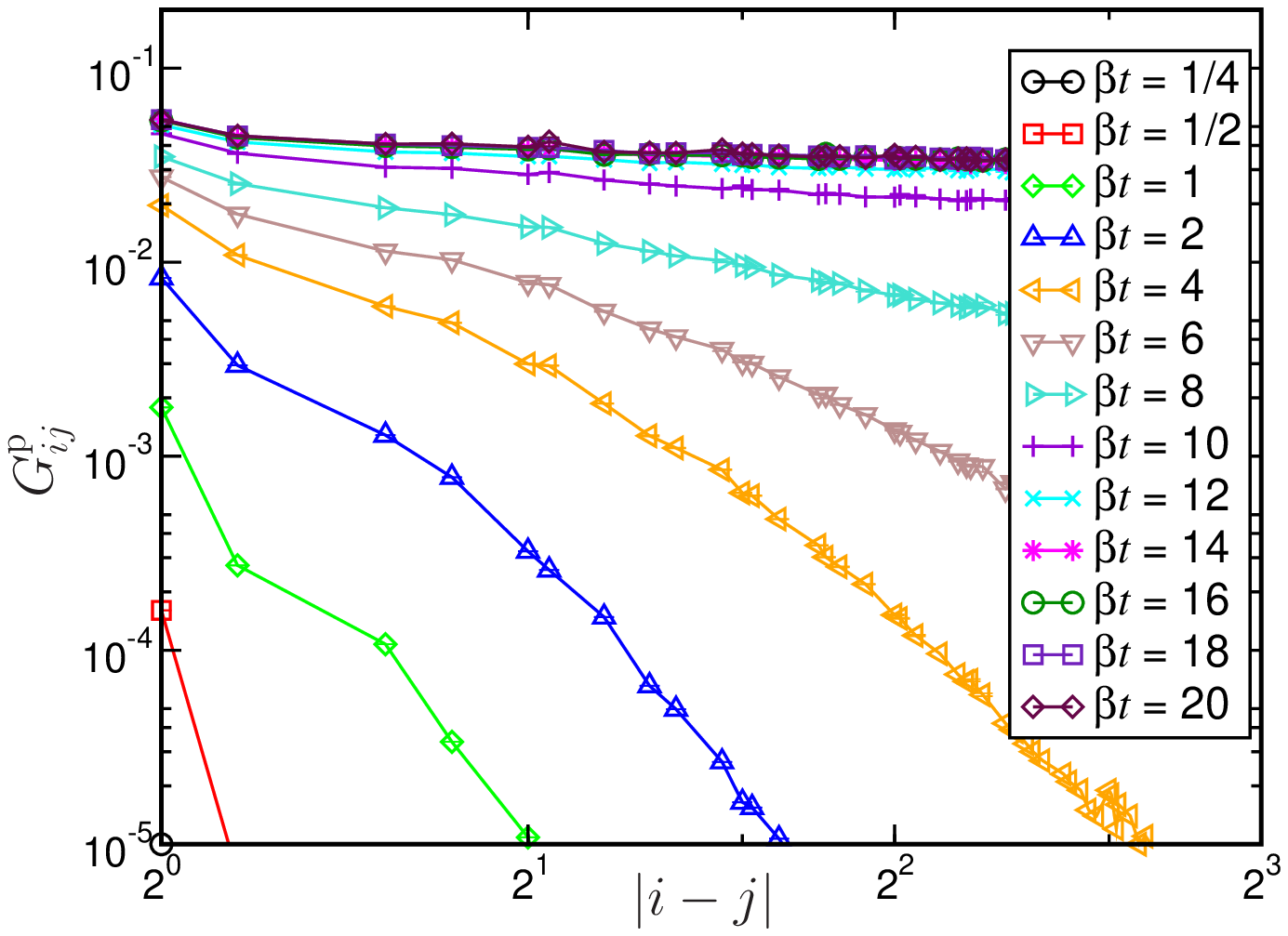} \caption{(Color online) Evolution
of the pair Green's function as a function of distance for different
temperatures. The total average fermionic density is set at $\rho=1.5$, the interaction strength at $U=3t$ and the lattice size is $L=12$. The vertical axes are plotted in logarithmic scale while the horizontal axes
are plotted with linear (top) and logarithmic (bottom) scales. For large site separation $|i-j|$, we observe a transition from an exponential decay (linear behavior in the log-linear plot) at high temperature to a weak algebraic decay (linear behavior in the log-log plot) at low temperature. This is the signature of the BKT transition where the system leaves the disordered phase to enter a phase with quasi-long-range order as the temperature is lowered. However, due to limited system size, the weak algebraic decay of the pair Green's function is difficult to infer unambiguously.  \label{fig:Green}} \end{figure}

A nice feature of the strongly-interacting regime is the existence of two
very different energy scales. One corresponds to the formation of tightly-bound 
pairs (molecules) and is typically of the order of $U$ itself. The second 
corresponds to the emergence of phase coherence between these pairs and
is of the order of the hopping parameter for pairs, typically $t^2/U$.\cite{spalek2007} 
These two energy scales are clearly
identified by comparing the evolution of $P_{\rm s}$ and $\rho_{\rm
p}$ when the temperature is varied, see Fig.~\ref{fig:Ps_and_rhop_vs_beta}. 
We thus can conclude that, even at $U/t=3$, we observe the formation of pairs
before the emergence of phase coherence. To investigate this
phenomenon further, we show in Fig.~\ref{fig:Green} the pair Green's function
(\ref{eqn:pairgreen}) as a function of distance for different
temperatures. There is a range of temperatures ($ 0.1 < \beta t< 5$)
where the pair Green's function is clearly decreasing exponentially
with distance (up to some boundary effects). This means that no phase coherence is achieved and 
the system is in a disordered regime. In other words, the
corresponding temperatures are above the BKT transition temperature $T_c$. For this same
temperature range, $\rho_{\rm p}$ has already reached its zero-temperature limit (Fig.~\ref{fig:Ps_and_rhop_vs_beta}). This is a clear evidence for the existence of preformed pairs which 
will eventually develop quasi-long-range phase coherence at a much lower temperature. For temperatures $T<T_c$, the Green's function should decay algebraically with distance with an exponent $\eta=T/(4T_c)$.\cite{paiva2004} For $\beta t\geq 10$, the pair Green's function behavior is consistent with a power-law decay, but it
is difficult to extract the corresponding exponent due to finite-size effects.

\section{Conclusion} We have studied the Hubbard model on a honeycomb
lattice with attractive interactions. At half-filling, building up upon previous existing studies, we have used the
mapping onto the FRHM to show that there is a
quantum phase transition at $T=0$ between a disordered phase and a DW-SF phase exhibiting crystalline as well as superfluid orders. The critical interaction strength at which this QPT takes place is accurately bounded by $5.0\leq U_c/t \leq 5.1$. We have also shown that, before the
transition, the system is semi-metallic and that the interactions do
not markedly change the nature of this phase. Away from half-filling,
within our numerical accuracy, the system seems to become superfluid,
even for arbitrary small values of the doping. We have elucidated the
presence of the BCS-BEC crossover by looking at several quantities,
especially the one-particle density of states. We have clearly evidenced,
for strong enough interactions, the existence of two different energy
scales, one for the formation of the pairs and one for the emergence
of phase coherence (the BKT transition), which is typical of the
strongly interacting regime.

For weak interactions, both at and away from half-filling, we have
observed that the spectral function $A(\omega)$ is qualitatively the
same as in the non-interacting case. Only the states close to the
Fermi level are affected by those weak interactions. As there are no
available states in the half-filled case close to the Fermi level, the
interactions hardly play a role and the system remains a semi-metal
(at half-filling) up to $U=5t$. It is only when the interactions are
strong enough to destabilize the Fermi sea and form tightly-bound
pairs that the system enters a different phase. In this case, the description in terms of individual fermions and plane-wave states is no longer relevant.

We further observe that the BCS and the semi-metal regimes are two phases sharing some common features. Indeed, in both phases, interactions are not strong enough to substantially modify the Fermi sea structure except around the Fermi level. This is reflected in the fact that the Dirac dip in $A(\omega)$ is always clearly visible in these cases. By the same token, the molecular superfluid phase (BEC) and the DW-SF have in common that the description in term of individual fermions is meaningless. Indeed, for both phases, the fermionic excitations are gaped and the Dirac dip in $A(\omega)$ has disappeared. Close to half-filling, we then observe the BCS-BEC crossover to happen for interaction strengths close to the value of the QPT at half-filling, {\it i.e.} $U=5t$.

\begin{acknowledgments} KL acknowledges support from the French
Merlion-PhD program (CNOUS 20074539). This work has also been
supported by the the France-Singapore Merlion program (SpinCold
2.02.07) and the CNRS PICS 4159 (France). Centre for Quantum Technologies is a Research Centre of
Excellence funded by the Ministry of Education and the National Research
Foundation of Singapore.  The work of RTS was supported under ARO
Award W911NF0710576 with funds from the DARPA OLE Program.  GGB is
supported by the CNRS (France) PICS 3659. We thank W. Zevon for useful
input.

\end{acknowledgments}

\bibliography{honeyV6}
\end{document}